\documentclass[%
 reprint,
superscriptaddress,
showpacs,
 amsmath,amssymb,
 aps,
prb,
]{revtex4-1}

\usepackage{graphicx}
\usepackage{dcolumn}
\usepackage{bm}
\usepackage{braket}

\graphicspath{{figures/}}

\begin{document}

\preprint{APS/123-QED}

\title{Bound state properties of \textit{ABC}-stacked trilayer graphene quantum dots}%

\author{Haonan Xiong}%
\thanks{These authors contribute equally to this work.}
\author{Wentao Jiang}%
\thanks{These authors contribute equally to this work.}
 \affiliation{%
Center for Quantum Information, IIIS, Tsinghua University, Beijing 100084, China
}%
 \affiliation{%
Department of Physics, Tsinghua University, Beijing 100084, China
}%
\author{Yipu Song}%
 \affiliation{%
Center for Quantum Information, IIIS, Tsinghua University, Beijing 100084, China
}%
\author{Luming Duan}%
\email{lmduan@umich.edu}
\affiliation{%
Center for Quantum Information, IIIS, Tsinghua University, Beijing 100084, China
}%
\affiliation{%
Department of Physics, University of Michigan, Ann Arbor, Michigan 48109, USA
}%

\date{\today}

\begin{abstract}
The few-layer graphene quantum dot provides a promising platform for quantum computing with both spin and valley degrees of freedom. Gate-defined quantum dots in particular can avoid noise from edge disorders. In connection with the recent experimental efforts [Y. Song \textit{et al.}, Nano Lett. \textbf{16}, 6245 (2016)], we investigate the bound state properties of trilayer graphene (TLG) quantum dots (QDs) through numerical simulations. We show that the valley degeneracy can be lifted by breaking the time reversal symmetry through the application of a perpendicular magnetic field. The spectrum under such a potential exhibits a transition from one group of Landau levels to the other group, which can be understood analytically through perturbation theory. Our results provide insight to the transport property of TLG QDs, with possible applications to study of spin qubits and valleytronics in TLG QDs.
\end{abstract}

\maketitle

\section{\label{sec:intro}introduction} 

Owing to its extraordinary electronic properties\cite{CastroElecPropOfG} and long coherence times,\cite{Probing-relax-time-in-GQD} graphene has received significant attention as a promising candidate for realization of quantum computing. Quantum dots (QDs) confined in graphene could be an ideal host for spin qubits.\cite{SpinQBInGQD,Quantum-dots-and-spin-qubits-in-graphene}  However, the electrostatic confinement of massless charge carriers has remained challenging due to the Klein tunneling and the absence of a gap in the spectrum.\cite{Chiral-tunnelling-and-the-Klein-paradox-in-graphene,SpinQBInGQD} So far, graphene QDs have been extensively investigated based on graphene nanoribbons and etched nanostructures, however, edge and substrate-induced disorder severely limits functionality of the device.\cite{elecStatesAndLL,ObservationOfElecHolePuddles,ElecHoleCrossover, QDBehaviorInGN} To avoid noise from edge disorders in graphene nanoribbons,\cite{gate-defined-graphene-double-quantum-dot} it is desirable to explore gate-defined graphene QDs.\cite{GateDefinedBLGQD} Few-layer graphene (FLG) is the only known material to exhibit a band structure depending on stacking and electric fields.\cite{stackingOrderDep} By breaking the layer inversion symmetry in AB-stacked bilayer or ABC-stacked trilayer graphene, an external perpendicular electric field can open an energy gap by local electrostatic gating.\cite{AsymmetryGapInBLG,SpinsInFEQD,EnergySpectrumOfABCTLG} Gate-defined and gate-controlled QDs have been demonstrated in bilayer\cite{GateDefinedBLGQD} and ABC-stacked trilayer graphene.\cite{Coulomb-Oscillations-in-a-Gate-Controlled-FLGQD} 

In order to design gate configuration in gate-defined QDs, numerical simulation is required to provide guidance. Landau level spectrum has been theoretically investigated in single and bilayer QDs.\cite{AnalyticModelOfGrapheneQD,BoundStateAndBFieldInducedVS} The results show that the valley degeneracy is broken by a magnetic field applied perpendicular to the graphene plane. M. Zarenia \textit{et al.} studied electron-electron interactions in BLG QDs under a parabolic potential by numerically solving the Schr\"{o}dinger equation.\cite{EEInteractionInBLG} Compared with monolayer and bilayer graphene, trilayer graphene has more complex interlayer interactions resulting in richer electronic structure.\cite{AvetisyanEFieldControl,stackingOrderDep} Several recent theoretical works have studied the Landau level spectrum of ABA- and ABC-stacked trilayer graphene.\cite{LLOfABAABCTLG,LLsInAsymmetricTLG} Numerical simulations reveal that six cubic bands of ABC TLG lead to three groups of Landau levels (LLs) with intergroup and intragroup LL anticrossings.\cite{EnergySpectrumOfABCTLG} TLG QDs with infinite-mass
boundary conditions have been studied recently.\cite{PhysRevB.94.165423} However, energy level spectrum and bound state properties have not been fully investigated yet in TLG QDs under a finite step potential well. In this paper, we present study of eigenspectrum in TLG QDs simulated with a finite step potential well using a general and analytic method. Our results show that the valley degeneracy of bound state levels can be lifted by a perpendicular magnetic field that breaks the time reversal symmetry, enabling possible control of spin qubits and valley-degrees of freedom in TLG QDs. Transition of energy levels between different groups of LLs can be identified from the calculation results and explained from the perturbation theory, which provides a guideline to identify the best parameter regime for designs of TLG QDs as qubits.

The paper is organized as follows. In Sec.~\ref{sec:bound_states} we introduce the analytic method to obtain the bound states in TLG QDs. The validity of our method is verified by matching the LLs in particular cases of homogeneous electrostatic potentials with results reported recently, and further confirmed by predicting the level transition between different groups of LLs from the perturbation analysis. Section~\ref{sec:results} includes the main results which identify the breaking of valley degeneracy and transition between different groups of LLs. We further analyze the results in Sec.~\ref{sec:discussions}, and discuss the best conditions to confine QDs.


\section{\label{sec:bound_states}bound states in ABC-stacked TLG} 
	
	Trilayer graphene has two different kinds of stacking, the HOPG stacking(ABA) and the rhombohedral stacking(ABC).\cite{elecStatesAndLL} To open a band gap by applying a perpendicular electric field,\cite{elecStatesAndLL,LLsInAsymmetricTLG} we focus our consideration on the ABC-stacked TLG in this paper. 

	\begin{figure}
	\includegraphics[width=2.5in]{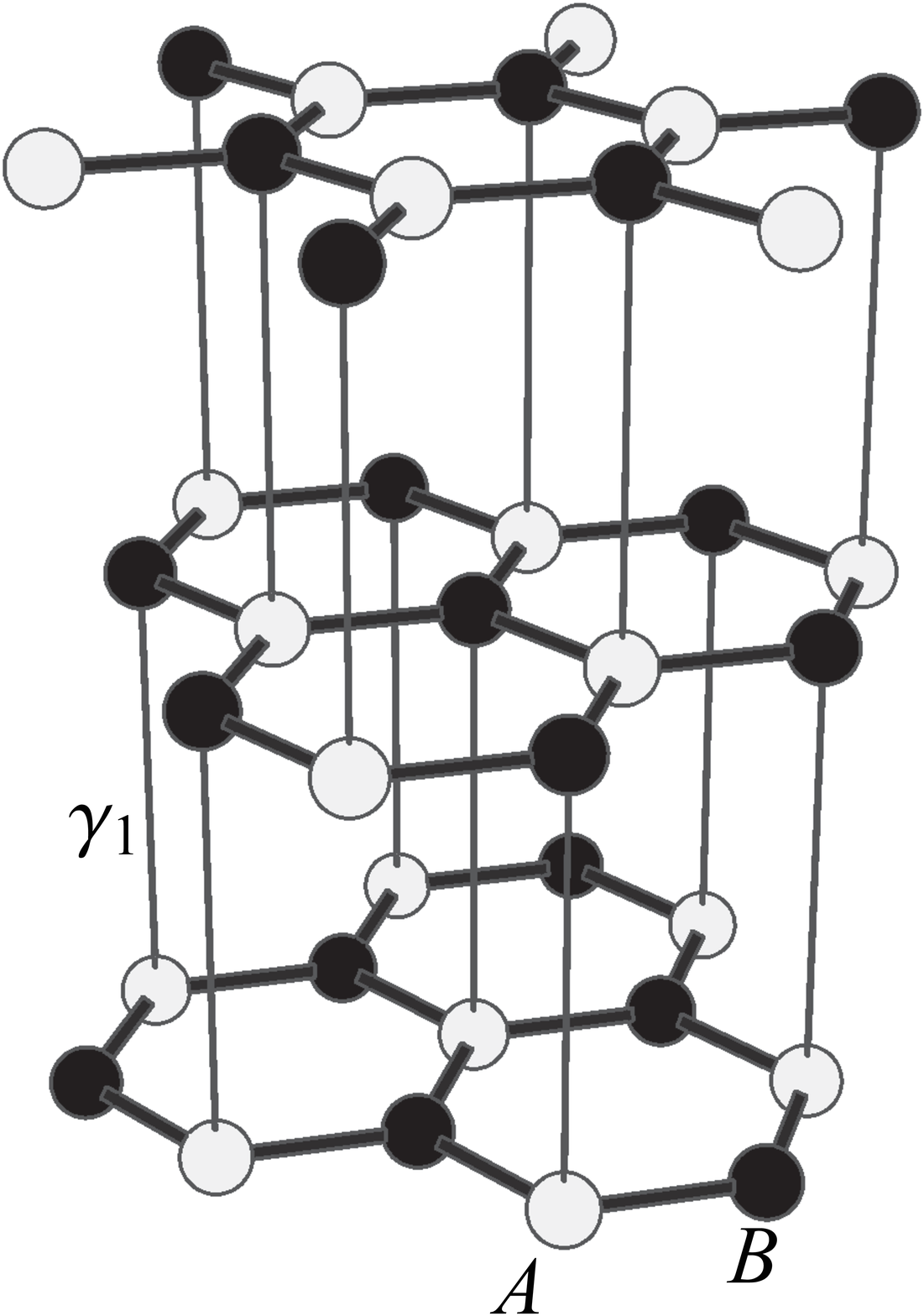}
	\caption{\label{fig:ABC_TLG}Atomic structure of ABC-stacked trilayer graphene. Intralayer hoppings are shown as black thick line and interlayer hoppings $ \gamma_{1} $ are shown as thinner gray line. Other intralayer and interlayer hoppings are neglected for simplicity. Site $ A$ (white) and site $B $ (black solid) are labeled in the figure on the bottom layer.}
	\end{figure}

	The atomic structure of ABC-stacked TLG is shown in Fig.~\ref{fig:ABC_TLG}. The nearest distance between adjacent carbon atoms is $a = 1.42 $\AA{} and the interlayer distance is $c_{0}=3.35 $\AA{}.\cite{AvetisyanEFieldControl} The different sublattices $A$ and $B$ are represented by white and black solid balls, respectively.
	
	To obtain the energy spectrum, we first generalize the analytic method for bound states in single layer and bilayer graphene by applying the step potential well to the ABC-stacked TLG. The validity of this method is tested by calculating the degenerate case (with homogeneous electrostatic potentials) and comparing the results with LLs obtained previously by different methods.\cite{LLOfABAABCTLG,LLsInAsymmetricTLG} Transition of bound state levels between two groups of LLs can be observed in the spectrum. The range where the transition occurs is consistent with the prediction from analysis based on perturbation theory.

	\subsection{\label{sub:analytic_solution}Analytic solution for bound states in ABC-stacked trilayer graphene QD} 

	We begin with the effective low-energy Hamiltonian of ABC-stacked TLG around the K point. Under the basis $( \psi_{A1},\psi_{B1},\psi_{A2},\psi_{B2},\psi_{A3},\psi_{B3}) $, where the components are envelope functions on different sublattices and different layers, the Hamiltonian is\cite{elecStatesAndLL,LLOfABAABCTLG}
	\begin{subequations}
	\begin{eqnarray}		
	\label{eqt:ham_p}	
		H &=& \left( \begin{array}{ccc}
			H_{p} & \Gamma & 0\\
			\Gamma^{\dagger} & H_{p} & \Gamma\\
			0 & \Gamma^{\dagger} & H_{p}
			\end{array} \right),\\
		H_{p} &=&  v_{F}\left( \begin{array}{cc}
			0 & p_{-}\\
			p_{+} & 0
			\end{array} \right), \Gamma = \left( \begin{array}{cc}
			0 & 0\\
			\gamma_{1} & 0
			\end{array} \right),
	\end{eqnarray}
	\end{subequations}
	where $p_{\pm}=p_{x}\pm i p_{y} $, $\bm{p}= (p_{x},p_{y})$ is the two-dimensional momentum operator.  $v_{\text{F} }  $ is the Fermi velocity of the monolayer graphene. For simplicity $\hbar=v_{\text{F}}=1 $. We only consider the nearest interlayer hopping  $\gamma_{1}=0.4$ eV. A homogeneous magnetic field $B$ is perpendicular to the TLG plane, which is included by the replacement $\bm{p}\rightarrow \bm{\pi}= \bm{p}+e\bm{A}(r) $. We denote the Hamiltonian after the replacement by $H_{\bm{\pi}} $.
	
	\begin{figure}
	\includegraphics[width=2.5in]{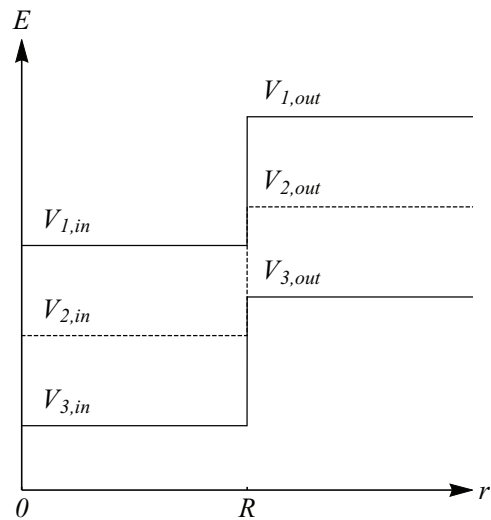}
	\caption{\label{fig:step_potential}Electrostatic potential used in the calculation. $R$ is the radius of the QD where the potential jump from $V_{i,\text{in}}$ to $ V_{i,\text{out}} $, leading to bound states. $R = 50$nm is adopted in this paper.}
	\end{figure}
	
	In order to simulate a quantum dot in trilayer graphene, we consider a piecewise constant electrostatic potential applied to the graphene. A schematic diagram of the potential is shown in Fig.~\ref{fig:step_potential}. The potential is fully characterized by six parameters $V_{i,\text{in}} $ and $V_{i,\text{out}} $, where $i=1,2,3 $ correspond to different graphene layers. This kind of potential is adopted as an approximation for gate-defined graphene QDs, where the potential barrier and gap are produced by top and back gates. The cases of more realistic potentials are briefly discussed in Sec.~\ref{sec:discussions}.

	The local gating effect is included in the potential term of the Hamiltonian, which is
	\begin{equation}
	\label{eqt:ham_V}
		H_{V}=diag[V_{1}(r),V_{1}(r),V_{2}(r),V_{2}(r),V_{3}(r),V_{3}(r)],
	\end{equation}
	where $r = \sqrt{x^{2} + y^{2}} $ is the distance from the center of the QD.

	Considering the rotation symmetry of the system, we choose the symmetric gauge $\bm{A} = B(-y,x,0)/2 $ and work in polar coordinates. The generalized momentum operator is found to be $\pi =\pi_{x}+i \pi_{y} = -ie^{i\theta} [\partial_{r} +i \partial_{\theta}/r -eBr/(2\hbar) ] $. Extending the total angular momentum operator in bilayer graphene,\cite{EEInteractionInBLG} the total angular momentum operator for a symmetric TLG system can be given as:
	\begin{equation}
	\label{eqt:J_z}
		J_{z}=L_{z}- \left[ \frac{\hbar}{2} \left( \begin{array}{ccc}
				I & 0 & 0\\
				0 & 3I & 0\\
				0 & 0 & 5I\\
				\end{array} \right)-\frac{\hbar}{2}\left( \begin{array}{ccc}
				\sigma_{z} & 0 & 0\\
				0 & \sigma_{z} & 0\\
				0 & 0 &\sigma_{z}
				\end{array} \right) \right],
	\end{equation}
	where $L_z=-i \partial_{\theta} $ is the orbital angular momentum which no longer commutes with the total Hamiltonian, $I$ is the $ 2\times 2$ identity matrix, $\sigma_{z} $ is the Pauli $ z $ matrix. $ J_{z} $ commutes with the Hamiltonian and the corresponding eigenstates of both operators are the six-component spinors:
	\begin{eqnarray}
	\label{eqt:eigenstate_psi}
		\Psi_{m}(r,\theta) &= e^{im\theta}\bm{(}\phi_{A1}(r),e^{i\theta}\phi_{B1}(r),e^{i\theta}\phi_{A2}(r),\notag\\
		&e^{2i\theta}\phi_{B2}(r),e^{2i\theta}\phi_{A3}(r),e^{3i\theta}\phi_{B3}(r)\bm{)}^T/\sqrt r,
	\end{eqnarray}
	where $m $ is the angular quantum number, $\phi_{Ai}$ and $\phi_{Bi}(i=1,2,3)$ are the envelope functions for different sublattice sites of the three graphene layers.

	The dimensionless coordinate $\xi=r/(\sqrt{2} l_{B}) $ can be defined to simplify the Hamiltonian, where $l_{B}=\sqrt{\hbar/(e|B|)} $ is the magnetic length. $s=\text{sgn} (B) $ refers to the direction of the magnetic field. The Hamiltonian acting on the six envelope functions $\psi(r)=(\phi_{A1},\phi_{B1},\phi_{A2},\phi_{B2},\phi_{A3},\phi_{B3})^T $ will be 
	\begin{subequations}
	\begin{eqnarray}		
	\label{eqt:ham_m}
		H_{m} &=& \left( \begin{array}{ccc}
				H_{1,m} & \Gamma & 0\\
				\Gamma^{\dagger} & H_{2,m} & \Gamma\\
				0 & \Gamma^{\dagger} & H_{3,m}
			\end{array} \right),\\
		H_{j,m} &=& \left( \begin{array}{cc}
				V_{j}(r) & \Delta_{B} \pi_{m+j-1}^{-}\\
				\Delta_{B} \pi_{m+j-1}^{+} & V_{j}(r)
			\end{array} \right),
	\end{eqnarray}
	\end{subequations}
	where $\Delta_{B} = 1/(\sqrt{2} l_{B}) $ is the magnetic energy and $\pi_{m}^{\pm} = -i[\partial_{\xi} \mp (m+1/2 )/\xi \mp s \xi] $ is the momentumlike operator acting on the components of the spinor $ \psi(r) $.

	The same functions in Ref.~\onlinecite{BoundStateAndBFieldInducedVS} can be adopted to further simplify the Hamiltonian:
	\begin{widetext}
	\begin{subequations}
	\label{eqts:phi}
	\begin{eqnarray}
	\label{eqt:phi_tot}
	    &\phi_{a}(m,s,\nu,\xi)=\left\{
				\begin{array}{lcl}
					\phi_{a}^{<}(m,s,\nu,\xi)	&	& {r\le R},\\
					\phi_{a}^{>}(m,s,\nu,\xi)	&	& {r>R},
				\end{array} \right.  \\
	\label{eqt:phi_in}
		&\phi_{a}^{<}(m,s,\nu,\xi)=\exp \left( - \frac{\xi^2}{2} \right) \xi^{|m+a|+1/2} M\left(\frac{|m+a|+1+ m-1-a}{2}s +\frac{\nu}{4},1+|m+a|, \xi^2 \right)/\Gamma(1+|m+a|),\\
	\label{eqt:phi_out}
		&\phi_{a}^{>}(m,s,\nu,\xi)=\exp \left( - \frac{\xi^2}{2} \right) \xi^{|m+a|+1/2} U\left(\frac{|m+a|+1+ m-1-a}{2}s +\frac{\nu}{4},1+|m+a|, \xi^2 \right),
	\end{eqnarray}		
	\end{subequations}
	\end{widetext}
	 where $M(a,b,z)$ and $U(a,b,z) $ are confluent hypergeometric functions.\cite{olver2010nist} $\phi_{a}^{<}$ is well-behaved at the origin and $\phi_{a}^{>}$ vanishes exponentially for $ \xi \rightarrow \infty$. $\nu$ is a parameter related to the eigenenergy. $a=0,1,2,3$ is relevant to layers and sublattices.

	Using the differentiation and recurrence relations of $M$ and $U$, the following relations can be obtained:
	\begin{subequations}
	\label{eqts:recur_of_phi}
	\begin{eqnarray}
		i \pi_{m+j-1}^{-} \phi_{j}=b_{2j-1}\phi_{j-1},\\{}
		i \pi_{m+j-1}^{+}\phi_{j-1}=b_{2j-1}\phi_{j},
	\end{eqnarray}		
	\end{subequations}
	with $j = 1,2,3$. $b_{1}\sim b_{6} $ only depend on $m,s$ and differ for different regions. For $r>R$,
	\begin{subequations}
	\label{eqts:b_i_out}
	\begin{eqnarray}
		b_{2j-1} &= -\left[ (1-s) +\left( \nu/4- j \right)(1+s) \right],\\
		b_{2j} &= -\left[ (1+s) +\left( \nu/4+ j \right)(1-s) \right].
	\end{eqnarray}		
	\end{subequations}
	For $r\le R$, the various values of $b_{i} $ are more complex and are shown in Table~\ref{tab:b_i_in}. 
	\begin{table}
		\caption{\label{tab:b_i_in}Values of $b_{i} $ for $r\le R$. }
		\begin{ruledtabular}
		\begin{tabular}{ccccc} 
			$b$ & $m \ge 0$ & $m=-1$ & $m=-2$ & $ m \le -3 $ \\
			\hline
			$b_{1} $ & $ 2 $ & $\nu/2-2s$ & $\nu/2-2s$ & $\nu/2-2s$ \\
			$b_{2} $ & $ \nu/2-2s $ & $2$ & $2$ & $2$ \\
			$b_{3} $ & $2$ & $2$ & $\nu/2-4s$ & $\nu/2-4s$ \\
			$b_{4} $ & $\nu/2-4s$ & $\nu/2-4s$ & $2$ & $2$ \\
			$b_{5} $ & $2$ & $2$ & $2$ & $\nu/2-6s$ \\
			$b_{6} $ & $\nu/2-6s$ & $\nu/2-6s$ & $\nu/2-6s$ & $2$ \\	
		\end{tabular}
		\end{ruledtabular}
	\end{table}
	Utilizing the properties of $\phi_{a} $ as in Eqs.~(\ref{eqts:recur_of_phi}), the Hamiltonian $H_{m} $ in Eq.~(\ref{eqt:ham_m}) can be simplified to a numeric matrix by rewriting $\psi(r)$ as
	\begin{equation}
	\label{eqt:psi_1_rewrite}
	\psi=(c_{A1} \phi_{0},c_{B1}\phi_{1},c_{A2}\phi_{1},c_{B2}\phi_{2},c_{A3}\phi_{2},c_{B3}\phi_{3})^{T}. 
	\end{equation}
	Notice that $\psi^{<}$ and $\psi^{>} $ are in different forms for $r \le R $ and $r>R $ respectively. Under this form of $\psi(r) $, the resulting Hamiltonian turns to be
	\begin{subequations}
	\begin{eqnarray}	
	\label{eqt:ham_num}	
		H_{b} &=& \left( \begin{array}{ccc}
				H_{1,b} & \Gamma & 0\\
				\Gamma^{\dagger} & H_{2,b} & \Gamma\\
				0 & \Gamma^{\dagger} & H_{3,b}\\
			\end{array} \right),\\
		H_{j,b} &=& \left( \begin{array}{cc}
				V_{j}(r) & -i \Delta_{B} b_{2j-1}\\
				-i \Delta_{B} b_{2j} & V_{j}(r) \\
			\end{array} \right),
	\end{eqnarray}
	\end{subequations}
	which is different for $r\le R $ and $r>R $. The Hamiltonian $H_{b} $ acts on numeric six-component spinor
	\begin{equation}
	\label{eqt:psi_2}
		\widetilde{\psi}=(c_{A1} ,c_{B1},c_{A2},c_{B2},c_{A3},c_{B3})^{T}.
	\end{equation}

	There are two equations $\text{det}(H_{b}-E)=0 $ for the eigenvalue problem, one for $r\le R $ and the other for $ r>R$. From Eqs.~(\ref{eqts:b_i_out}) and Table~\ref{tab:b_i_in}, $\nu$ always appears three times in the Hamiltonian in Eq.~(\ref{eqt:ham_num}). Hence $\nu_{i}^{<} $ and $\nu_{i}^{>} $ can be solved as lengthy algebraic expressions of $E$, where $i=1,2,3 $ distinguish the three solutions and the superscript refers to the two regions. For every single eigenenergy $E$, there are three corresponding eigenstates for both $ {\widetilde{\psi}}^{<}_{\nu_{i}} $ and ${\widetilde{\psi}}^{>}_{\nu_{i}} $. In total, $\widetilde{\psi} $ combined with Eq.~(\ref{eqt:eigenstate_psi}), Eqs.~(\ref{eqts:phi}) and Eq.~(\ref{eqt:psi_1_rewrite}) gives us three possible eigenstates $ \Psi_{m,\nu_{i}}^{<}$ for region $ r \le R $ and three states $ \Psi_{m,\nu_{i}}^{>}$ for region $ r > R $, where the suffix $ \nu_{i} $ represents the value of parameter $\nu $. The final eigenstates are combination of the three $ \Psi_{m,\nu_{i}} $ for both regions, which should have identical value at the point $r=R $, i.e.,
	\begin{equation}
		\label{eqt:connect_condition}
		\sum_{i=1}^{3} c_{i}^{<} \left. \Psi_{m,\nu_{i}}^{<}\right|_{r=R} = \sum_{i=1}^{3} c_{i}^{>} \left. \Psi_{m,\nu_{i}}^{>} \right|_{r=R},
	\end{equation}
	where $c_{i}^{<} $ and $c_{i}^{>} $ are combination coefficients. Notice that all $ \Psi_{m,\nu_{i}}^{<}$ and $ \Psi_{m,\nu_{i}}^{>}$ are six-component spinors, so Eq.~(\ref{eqt:connect_condition}) is a set of six equations for six unknown $c_{i}^{<} $ and $c_{i}^{>} $. The pre-existing condition of the solutions of $c_{i}^{<} $ and $c_{i}^{>} $ gives the equation with solvable $E$, namely,
	\begin{equation}
	\label{eqt:det_for_E}
		\text{det}\left.\left(\psi_{\nu_{1}}^{<},\psi_{\nu_{2}}^{<},\psi_{\nu_{3}}^{<},\psi_{\nu_{1}}^{>},\psi_{\nu_{2}}^{>},\psi_{\nu_{3}}^{>}\right)\right|_{r=R}=0,		
	\end{equation}
	where $E $ is implicitly included in $\nu_{i}^{<} $ and $\nu_{i}^{>} $. $\psi$, as defined in Eq.~(\ref{eqt:psi_1_rewrite}), is used instead of $\Psi_{m}$, as the angular components are the same in every row in this determinant and can be cancelled out. Subscript $ \nu_{i} $ distinguishes different values of $\nu $ used in $\psi$.

	Equation~(\ref{eqt:det_for_E}) is the analytic equation to solve the bound state energy levels of the QD in ABC-stacked trilayer graphene, which can be obtained numerically.
	

	\subsection{\label{sub:comparison}Comparison with LLs}

\begin{figure}
	\includegraphics[width=3.4in]{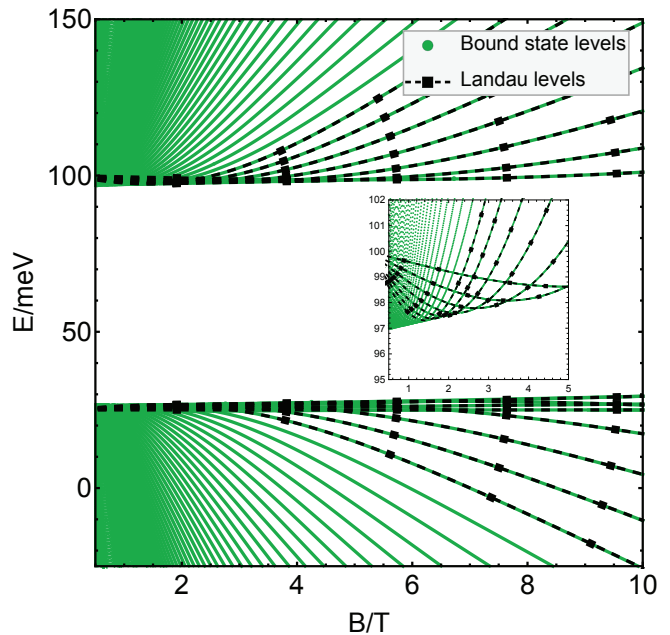}
	\caption{\label{fig:homo_pot_ELs}Typical low-lying energy levels versus magnetic fields for homogeneously biased ABC-stacked trilayer graphene. Green dots: levels obtained by the analytical method for homogeneous potential as a degenerate case. Arbitrary $m$ and $R$ gives the same results. Black dashed line: Landau levels obtained by method introduced in Appendix.~\ref{sec:energy_levels_in_homogeneously_biased_tlg}. Potentials $ V_{1} = 100 $ meV, $ V_{2} = 50 $ meV, and $V_{3} = 25$ meV are adopted. }
	\end{figure}

	To test the validity of our method, we apply the method to the degenerate case, i.e., $V_{i,\text{in}} = V_{i,\text{out}} = V_{i} $ for all $ i=1,2,3 $ and compare the results with the simulation results reported recently, where the energy spectrum should be reduced to LLs in homogeneous electric and magnetic fields. As a typical case, $ V_{1} = 100 $ meV, $ V_{2} = 50 $ meV, and $V_{3} = 25$ meV are used and the resulting levels are shown in Fig.~\ref{fig:homo_pot_ELs} as a function of the magnetic field. Green dots represent energy levels obtained by the analytic method introduced in Sec.~\ref{sub:analytic_solution}. The first $7$ LLs are included as black dashed lines, which are obtained by method introduced in Appendix.~\ref{sec:energy_levels_in_homogeneously_biased_tlg}. A zoom-in plot of energy range near the band edge is shown in the inset. Reversion of LL order can be observed in the region of relatively weak magnetic fields. As can be seen from Fig.~\ref{fig:homo_pot_ELs}, bound state levels, in the case of degenerate potential, agree perfectly with LLs.

	The potentials adopted here are the same as in Ref.~\onlinecite{LLsInAsymmetricTLG}, where the LLs for asymmetric TLG are obtained by solving the coupled Hamiltonian equations. An opening of a band gap can be observed. An order reversal of the LLs also appears for weak magnetic fields. Results from our calculation perfectly agree with results obtained by directly solving the coupled differential equations.\cite{LLsInAsymmetricTLG} The two groups of LLs start from $V_{1} $ and $V_{2} $ respectively, which also set the range of the band gap.\cite{LLsInAsymmetricTLG}

	For a step potential in general, the bound state levels go to LLs only under strong magnetic fields when $ l_{B} \ll R $. Eigenstates can be separated into three sets: mainly inside the QD, mainly outside the QD or sitting across the potential step, which depends on the magnetic field. For the first two sets, the states experience a nearly homogeneous electrostatic potential and are very close to states corresponding to LLs, and so are the bound state levels. Transition of levels from outside LLs to inside LLs can be observed, when the magnetic field increases and affects the bound states to shrink from outside, crossing the step and being into the QD. The above arguments can be formulated by the perturbation theory on the potential and are discussed in detail in Appendix.~\ref{sec:expansionByLL}. For simplicity, LL states in the following context refer to the eigenfunctions corresponding to Landau levels. The perturbation theory is based on LL states in polar coordinates, which can give an estimate on the transition range. The transition range here is defined to be the range of the magnetic field $B$ under which the energy levels transit from one group of Landau levels to another group as $B$ increases. We present the result given by the perturbation method here to further demonstrate the validity of the analytic solution and the perturbation method itself.

\begin{figure}
		\includegraphics[width=3.4in]{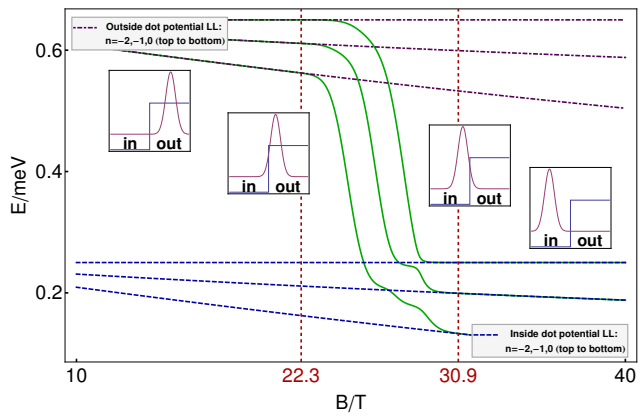}
		\caption{\label{fig:LLEsinglem} Transition between two groups of LLs for bound state levels with $ m = -800$ and $ n=0,-1,-2$. Upper three dot-dashed lines are LLs corresponding to the outside potential for $n=0,-1,-2$. Lower three dashed horizontal lines are LLs corresponding to the inside potential for $n=0,-1,-2$. The solid green lines are bound state levels under the step potential for $m=-800$. Two dashed vertical red lines indicate the transition range for $n=0, m=-800$ obtained by the perturbative method. The four insets represent the spatial relationships of the wave function and the dot under four different magnetic fields }
		\end{figure}

\begin{figure}
		\includegraphics[width=3.4in]{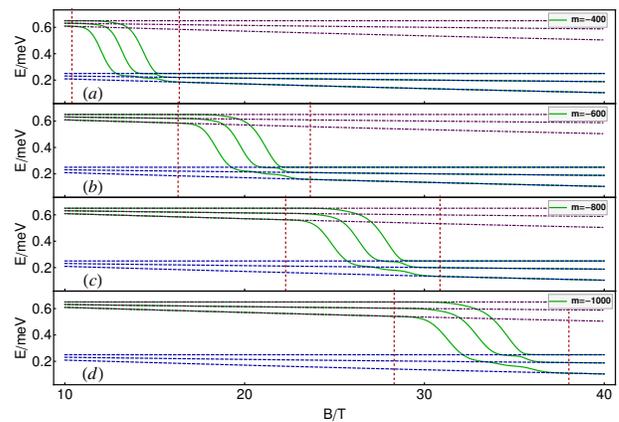}
		\caption{\label{fig:LLEmultim} Transition between two groups of LLs for bound state levels with different $m$'s, all consistent with transition range obtained by the perturbative method. $m=-400, -600, -800, -1000$ are adopted in the sub figures (a) to (d), respectively.}
		\end{figure}

		A step potential described by $ V_{1,\text{in}}=0.25 \mathrm{ meV}, V_{2,\text{in}} = 0  \mathrm{ meV}, V_{3,\text{in}} = -0.25 \mathrm{ meV} $ and $V_{i,\text{out}} = V_{i,\text{in}} + 0.4 \mathrm{ meV} $ is adopted here. The analytical method gives the energy spectrum. From the expansion perspective in the perturbation method, the comparison between the spatial extension of a LL state and the dot area gives the transition range of the bound state levels in Fig.~\ref{fig:LLEsinglem} and Fig.~\ref{fig:LLEmultim}.		
		
		Figure~\ref{fig:LLEsinglem} shows a zoomed area of the energy spectrum and picks out three energy levels with $m=-800$ and $n=0,-1,-2$ which displays the transition of bound states between different LLs in this condition. The transition range is given by two vertical red dashed lines. If the magnetic field is stronger than the upper limit of this range, the energy levels will be very close to LLs for $n=0, -1, -2$ with respect to the potential inside the dot, as the corresponding LL states are located mainly inside the dot. The same argument can be applied to the magnetic field weaker than the lower limit of this range. For the magnetic fields in this range, the energy level will exhibit a transitional behavior to connect the landau levels on the left and right side. The eigenfunctions are the superposition of LL states with different $n$'s, which is indicated by the anticrossings at the end of the transition. The transition ranges for more $m$'s are illustrated in Fig.~\ref{fig:LLEmultim}, indicating the validity of such expansion perspective and the analytic method in general for different $m$'s.

	\section{\label{sec:results}Results} 
	
	We show in Sec.~\ref{sub:comparison} that, for a homogeneously biased ABC-stacked TLG with layer potential $V_{1} $, $V_{2} $ and $V_{3} $, a bandgap ranging from $V_{1} $ to $V_{3} $ can be opened near zero energy. Two groups of LLs can be identified. One starts from the upper edge of the bandgap and bends upward. The other starts from the lower edge and bends downward.

	For the piecewise constant potential in Fig.~\ref{fig:step_potential} , it's straightforward to think that the energy levels of the system should be a combination of LLs from $V_{i,\text{in}} $ and those from $V_{i,\text{out}} $. It is expected that there are  four groups of levels near zero energy. Two of them show an energy gap from $V_{1,\text{in}} $ to $V_{3,\text{in}} $ and the other two show another gap from $V_{1,\text{out}} $ to $V_{3,\text{out}} $. According to whether the two gaps overlap or not, two different cases can be resolved with different overall potentials. To be explicit, if we have $V_{1,\text{in}} > V_{3,\text{out}} $, which is the case in Fig.~\ref{fig:step_potential}, the intersection of the two gaps gives a true gap in the energy levels. If $V_{1,\text{in}} < V_{3,\text{out}} $ instead, the LLs corresponding to $ V_{i,\text{out}} $ will have one group bending downward from $ V_{3,\text{out}} $, crossing with another group of LLs which are bending upward from $ V_{1,\text{in}} $. The true energy levels of bound states have a complex patterns for weak magnetic fields and converge to LLs for strong magnetic fields.

	In this section, we numerically solve the bound state energy levels from Eq.~(\ref{eqt:det_for_E}) for two different potentials, i.e., two sets of $ V_{i,\text{in}} $ and $ V_{i,\text{out}} $, one with finite intersection of inner and outer bandgap, resulting in an energy gap in the bound state levels, and the other with no intersection, giving a complex crossing feature in the evolution of energy levels versus the magnitude of the magnetic fields.

	The first set of $ V_{i,\text{in}} $ and $ V_{i,\text{out}} $ is adopted as follows
	\begin{subequations}
		\label{eqts:potential_overlap_gap}
	\begin{eqnarray}
		V_{1,\text{in}} &=&  \tau \frac{V}{2},\notag\\
		V_{2,\text{in}} &=& 0,\\
		V_{3,\text{in}} &=& - \tau \frac{V}{2}, \notag\\
		V_{i,\text{out}} &=& V_{i,\text{in}} + U,
	\end{eqnarray}		
	\end{subequations}
	where $\tau $ accounts for the valley degree of freedom and $V = 50 $ meV, $U = 40 $ meV are used for the calculation. $ V_{1} $ and $ V_{3} $ are exchanged for $ \tau=-1 $. The valley degree of freedom can be included by $\tau$ and will be briefly discussed in Appendix.~\ref{sec:valley_degree_of_freedom_in_tlg}.

	\begin{figure}
	\includegraphics[width=3.4in]{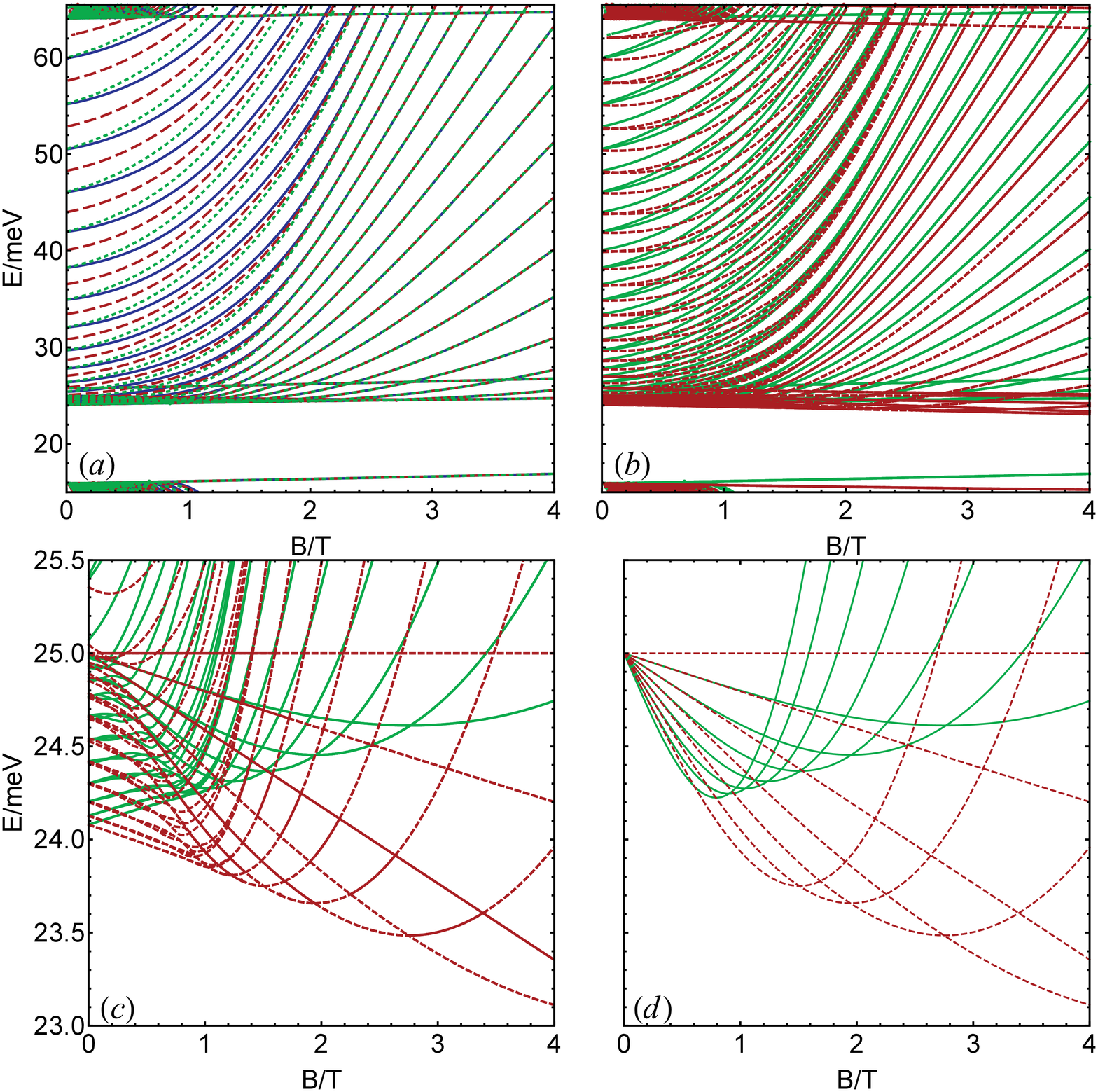}
	\caption{\label{fig:levels_potential1}Bound state levels of a trilayer QD. (a) $m=-1$(blue solid lines), 0(red dashed lines), 1(green dotted lines) for $\tau=1$. (b) Both $\tau=1$ (green solid lines) and $\tau=-1$ (red dashed lines) are included and $m=0,\pm 1$ for each value of $ \tau $. (c) Zoom-in of levels near the upper edge fo the gap, green solid lines for $\tau=1$ and red dashed lines for $\tau=-1$. (d) The first seven LLs corresponding to potential inside the QD, of the same energy range as in (c), green solid lines for $ \tau=1 $ and red dashed lines for $\tau=-1$. }
	\end{figure}

	The results of the bound state levels under the above potential are displayed in Fig.~\ref{fig:levels_potential1}, which shows the bound state energy levels with respect to the magnetic field. Figure~\ref{fig:levels_potential1}(a) includes levels for $m = -1,0,1 $ and $ \tau = 1 $. Degeneracy of levels for different $m$ values can be clearly observed at $B=0$. Levels with different $m$'s converge to the same LLs as long as they have the same LL index $n$ within the same group of levels. Figure~\ref{fig:levels_potential1}(b) shows the bound levels for both $ \tau=\pm 1 $ for the same range of $m$. Levels corresponding to different valleys are degenerate in the zero magnetic field, however, the degeneracy is lifted for the finite magnetic fields. An increase in level densities can be observed near the band edge, which agrees with the ``Mexican hat'' band structure of ABC-stacked trilayer graphene in the presence of a perpendicular electric field. \cite{band-structure-of-ABCTLG}

	A zoom-in plot of the bound state levels near the band edge is shown in Fig.~\ref{fig:levels_potential1}(c). Lift of valley degeneracy can be clearly observed as $B$ increases from zero to finite values. In the strong magnetic fields where $ l_{B}\ll R $, bound state levels tend to bulk LLs. The corresponded LLs are included in Fig.~\ref{fig:levels_potential1}(d) for $ \tau=\pm 1$, which is in perfect agreement with bound state levels in strong magnetic fields.

	The other set of $ V_{i,\text{in}} $ and $ V_{i,\text{out}} $ is used with $ V_{1,\text{in}} =  -8.6$ meV, $ V_{2,\text{in}} = -2.4 $ meV, $ V_{3,\text{in}} = -2.3 $ meV, $ V_{1,\text{out}} = -0.5 $ meV, $ V_{2,\text{out}} = 3.7 $ meV and $ V_{3,\text{out}} = 9.3 $ meV, where the ranges of the two energy gaps have no intersection.

	\begin{figure*}
	\includegraphics[width=7in]{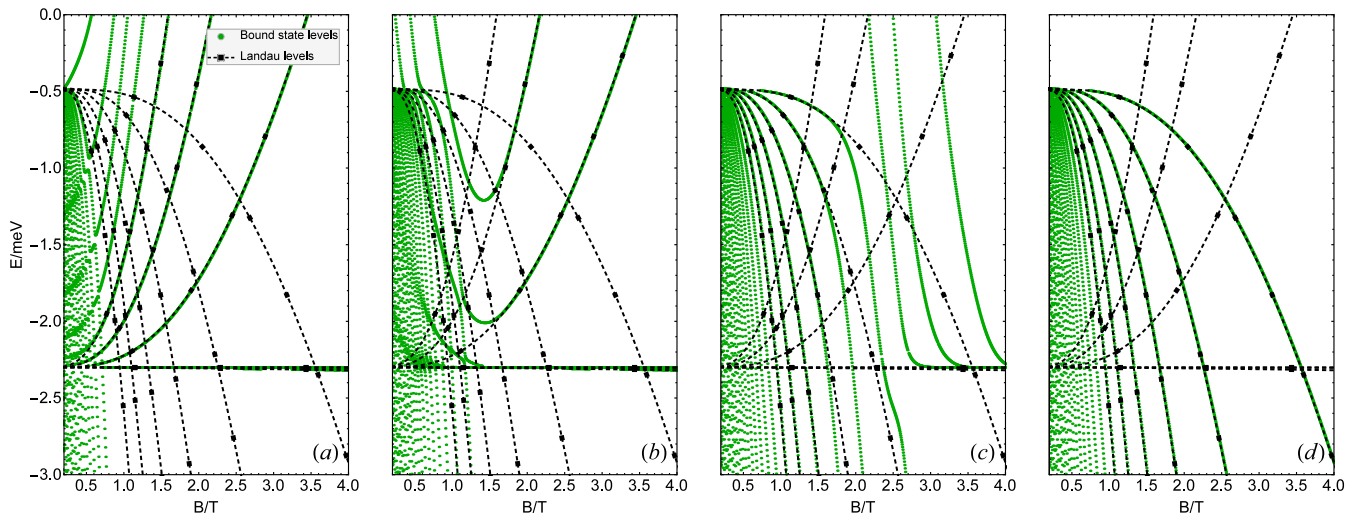}
	\caption{\label{fig:levels_potential2}Energy levels of a trilayer QD under one typical second-type potential. Different $ m $ values are used for different sub-figures. (a) $ m = 0$. (b) $m = -20 $. (c) $ m = -80$. (d) $ m = -200$. Bound state levels are shown in green dots. Black dashed lines are the first six LLs. Anticrossings can be observed and transition from inner LLs to outer LLs occurs at different magnetic field strength for different $m $ values.}
	\end{figure*}

	Figure~\ref{fig:levels_potential2} shows the results of energy levels as a function of $B$ for the second set of $ V_{i,\text{in}} $ and $ V_{i,\text{out}} $. The range of energy used in the plot is chosen so that the crossing feature is included. Two groups of LLs are also included, corresponding to homogeneously biased TLG under $ V_{i,\text{in}} $ and $ V_{i,\text{out}} $ respectively. The first six LLs are included as black dashed lines. One group of LLs originates from $ V_{3,\text{in}} = -2.3 $ meV and bends upward, which sets the upper edge of the energy gap corresponding to $ V_{i,\text{in}} $. The other group of LLs originates from $ V_{1,\text{out}} = -0.5 $ meV and bends downward, which sets the lower edge of the energy gap corresponding to $ V_{i,\text{out}} $. The bound state levels are plotted as green dots, and four different values of $ m $ are used in the four subfigures. The bound state levels show interesting and complex evolution patterns with respect to the magnetic field for different values of $ m $, exhibiting transitions between different groups of LLs.
		


\section{Discussion} 
\label{sec:discussions}

	In the previous calculation, the valley degeneracy can be lifted by a magnetic field, which is an essential step towards controlling valley degree of freedom in graphene. Previous work has shown that graphene ribbons can be used as a valley filter.\cite{Valley-filter-and-valley-valve-in-graphene,Valley-Contrasting-Physics-in-Graphene} However, it's hard to yield a graphene sheet with a deterministic edge shape. In our gate-defined QD with valley splitting, especially near the edge of the band gap, large energy difference between the two valley makes it possible to select states with definite valley, realizing a valley polarization for current passing through the QD.
	
    The valley splitting $\Delta_{K,K'}$  can be calculated as the energy difference between one valley-polarized ground state and the first excited states from the other valley,\cite{BoundStateAndBFieldInducedVS} which is about 1.6 meV from Fig.~\ref{fig:levels_potential1} on condition that $B=4$T. This is larger than Zeeman splitting $g\mu_{B}B \approx 0.47$ meV with $g\approx2$ (Ref.~\onlinecite{Spin-States-in-Graphene-Quantum-Dots}). As a consequence, the valley  freedom will not entangle with the spin freedom, resulting in longer coherence time of the spin qubit in TLG QDs.
    
	In addition to valleytronics, the energy spectrum and its expansion interpretation can provide useful insights for the quantum dots in the presence of a magnetic field. For the energy levels between the gap outside the dot, the corresponding states are localized inside the dot and well confined. So the Fermi energy within such energy range is appropriate for confining a quantum dot. However, for the energy levels outside this gap, even though there can be energy levels corresponding to bound states confined inside the dot, the energy levels for the LL states outside the dot intersect with other levels. In consequence, the electron will possibly escape from these degenerated LLs, making the dot ``leaky''. For a special case in which $V_{1,\text{in}}<V_{3,\text{out}}$, the leak of the electrons in the conduction band inside the dot, whose energies also lie within the valence band of the barrier, can be thought as Klein tunneling because in this case the electrons escape through the valence band with respect to the barrier potential.\cite{Andreev-Reflection-And-Klein-Tunneling-in-Graphene} From the theory of Klein tunneling, one can deduce that increasing the magnetic field would suppress the conductance.\cite{Andreev-Reflection-And-Klein-Tunneling-in-Graphene}

	As mentioned in Sec.~\ref{sec:results}, when the gap inside the dot and that outside the dot overlap with each other, a true gap emerges and forbids the tunneling of electrons. In this circumstance, the exhaustion of electrons or holes can be observed. Considering the potential in  Fig.~\ref{fig:step_potential}, if only $V_{1,\text{in}} $ lies within $V_{1,\text{out}} $ and $V_{3,\text{out}} $, which is the case illustrated in Fig.~\ref{fig:step_potential}, the exhaustion of electrons is permissible. If only $V_{3,\text{in}} $ lies within the outside gap, the exhaustion of holes can be obtained. If the entire inside gap is within the outside gap, a transition from electron QD to hole QD can show up in the transport measurement. Due to the existence of charging energy,\cite{spin-in-few-E-QD} the exhaustion phenomenon can only be observable if the gap is comparable or larger than charging energy. If the entire outside gap is within the inner gap, there will either be no current or relatively large current as the Fermi levels being tuned across the band gap of the inside potential.

	For the simple step potential in Fig.~\ref{fig:step_potential}, the eigenstates under magnetic fields can be taken as a combination and hybridization of the LL states corresponding to the potential inside and outside the dot. In real experiments, the potential profiles are more complex. To further predict the energy level tendency, extensions can be made from the simple step potential well to three types of more complex potentials: smooth edge potentials, nonaxial symmetric potentials and multistep potentials. The smooth edge potential  refers to the potential which has a smooth transition at the dot edge. For this kind of potentials, the transition ranges of energy levels mentioned in Sec.~\ref{sub:comparison} will be broader than those of the step potential, but there is no essential difference. The nonaxial symmetric potential has a dependence on $\theta$. Such nonaxial symmetric dot area will also broaden the transition ranges of the energy levels. This is because the distortion of the dot edge from a circle can be restricted by a ring if the distortion is not so large. Outside the ring, the potential is flat and possesses axial symmetry, therefore the ring can be taken as a transition area of the potential. Similar to the smooth edge potential's transition area, it causes hybridization of LL states under a wider magnetic field transition range. The difference from the smooth edge potential case is that, inside the transition ranges, the breaking of the rotational symmetry will mix the LL states belonging to different $m$'s together. The multistep potential can be described as 
\begin{equation}
	\label{eqt:multi-step_potential}
	V_{i}(r)=\left\{
		\begin{array}{lcl}
			V_{i1}	&	& {r \le R_{1}},\\
			V_{i2}	&	& {R_{1} \le r<R_{2}},\\
			\dots 	&	& \dots \\
			V_{in}  &   & {R_{n-1} \le r},
		\end{array}	\right.
\end{equation}
	where $ i=1,2,3 $. The energy spectrum under such potential can be foreseen as combination and hybridization of the Landau levels corresponding to each of the step. In this way, even though the rotational symmetry of the potential is retained, more degeneracy of $m$ will be lifted and more transition ranges will emerge. Further extension can be made on other 2D materials. Once the Landau levels and LL states are obtained, the bound states and energy spectrum tendency, under a dot potential and magnetic fields, can be predicted similarly as the situation for the TLG quantum dot.	If a more accurate prediction on the spectrum is needed, one may follow the perturbation process, calculate the integrals and solve the eigenvalue problem. An alternative way is to directly solve the differential equation numerically such as using the commercial software COMSOL, which can give similar results in our case.
	
	In order to increase the coherence time of TLG-QD-based spin qubits, all the degeneracy should be broken to prevent undesirable incoherent mixture of other degrees of freedom and avoid their entanglement with the spin.\cite{BoundStateAndBFieldInducedVS} In addition to valley degeneracy, the orbital degeneracy for different $m$'s should also be lifted off, which can  be obtained with the electrostatic potential. So the magnetic field should be inside transition range and the transition should be steep enough for a relatively large energy difference between the concerned orbital energy levels. For a step potential well in our case, because the LL states of larger $n$ have wider spatial extension, the transition ranges are also wider and not steep. As a result, the energy difference between different $m$'s will be relatively smaller comparing to those with lower $n$. So it is better to adjust the Fermi energy to the energy levels corresponding to lower $n$ and avoid the higher $n$ regime. For TLG, a typical band gap adopted in Fig.~\ref{fig:levels_potential2} is approximately $10$ meV (the gap outside the dot). In the cases like this, most high $n$ levels inside the gap correspond to the magnetic field strength smaller than 1T. So in this case, the magnetic field should be larger than 1T ,but it should not be too large at the same time. On the one hand the LLs for $n=0$ tend to shrink the gap, and on the other hand the levels of different $m$'s will get closer. The distribution of the energy levels near the band edge is determined by $R/l_{B}$ for the step potential well. This value should be neither too small to have a proper Zeeman splitting nor too large to keep levels with different angular momentum from being nearly degenerate again. The size of the dot will determine the steepness of the potential well and thus affect the energy difference between different angular momentum in the transition range.
	
	Since the wave function can be obtained analytically in the step potential well, it enables the estimation for the exchange interaction between two spin qubits of the graphene QDs. For a double dots system, the bound states in each dot can take similar forms and be related with a phase factor from gauge transformation. In this way, the exchange energy $J$ can be calculated as the energy difference of the singlet and the triplet.\cite{Coupled-Quantum-Dots-as-Quantum-Gates} Once the exchange coupling $J(t)$ is obtained, a SWAP operation can be realized by tuning $J(t)$ with the gate voltage and the magnetic field. The SWAP gate can be used to construct the XOR gate which is universal. For distant QDs, coupling can be further achieved via various architectures,\cite{delbecq2013photon,deng2015} enabling potential applications in quantum information processing.
	
	Other than completely numerical ways to find eigenfunctions, the expansion approach may provide a general method for real potentials in experimental situations. As discussed above, the ideal energy levels for TLG QD qubits should be within the gap outside the dot and the ideal magnetic field should be larger than 1T, thus an upper limit for $m$ and $n$ can be set in the perturbation process and this makes the expansion method possible for arbitrary potential shape.

\section{\label{sec:conclusions}conclusion} 

In summary, we have solved the bound state levels of ABC-stacked TLG QDs under a step potential well. Similar to the cases in single layer and bilayer graphenes,\cite{BoundStateAndBFieldInducedVS} breaking of valley degeneracy is observed under a homogeneous magnetic field. Transfer to LLs can be identified under strong magnetic fields. We test the validity of our method by calculating the degenerate case with homogeneous electrostatic potentials with $ V_{i,\text{in}} = V_{i,\text{out}} = V_{i} $. The results agree with LLs obtained by previous theoretical studies.\cite{LLOfABAABCTLG,LLsInAsymmetricTLG} Transition of bound state levels can be seen between two groups of LLs with increasing magnetic field strength. The range, in which the transition occurs, is consistent with the  prediction from a perturbative analysis. The step potential well can be distinguished into two cases, depending on whether the band gap of inner LLs and outer LLs overlap or not. For the first case, a true energy gap will occur in the overlapping energy range and relatively large energy deviation between two valleys can be observed near the band edge. We have discussed the consequences and potential applications of valley splitting. We also exploit possible generalization of step potential and the resulting pattern of bound state levels from the perturbative analysis. Optimal parameters for TLG QD qubits can be explored from our spectrum analysis. Our method also paves the way for prediction of exchange interaction between TLG QDs.
	

\begin{acknowledgments}
	This work was supported by the Ministry of Education of China through its grant to Tsinghua University.
\end{acknowledgments}

\appendix

	\section{\label{sec:energy_levels_in_homogeneously_biased_tlg}Energy levels in homogeneously biased TLG} 
	

	Under constant electrostatic potential, the TLG system can be described by the same $H_{\bm{p}} $ as in Eq.~(\ref{eqt:ham_p}) and also has the same substitution $\bm{p}\rightarrow \bm{\pi}= \bm{p}+e\bm{A}(r) $ for the homogeneous magnetic field. $H_{\bm{\pi}} $ can be diagonalized by first choosing the Landau gauge $\bm{A} (\bm{r} )=(0,Bx) $ and then adopting LL wave functions to simplify $H_{\bm{\pi}} $. After this procedure, $H_{\bm{\pi}} $ is reduced to a numeric matrix of which the eigenvalues can be easily solved.\cite{MOPropertiesOfMLG,LLOfABAABCTLG} We point out here that the method in Ref.~\onlinecite{LLOfABAABCTLG} can be directly adopted to the system under a homogeneous external electric field described by
	\begin{equation}
		\label{eqt:ham_Vhomo}
		H_{V}=diag[V_{1},V_{1},V_{2},V_{2},V_{3},V_{3}],
	\end{equation}
	where $V_{i} $'s are constants independent of space position. The energy levels of the system $ H = H_{\bm{p}}+H_{V} $ can be solved from the above method for given layer potentials $V_{i} $.

	
	\section{\label{sec:valley_degree_of_freedom_in_tlg}Valley degree of freedom in TLG}
	
	The valley degree of freedom is originally included in the low energy effective Hamiltonian in Eq.~(\ref{eqt:ham_p}) as $ p_{\tau,\pm} = \tau  p_{x}\pm i p_{y} $, where $ \tau = \pm 1 $ distinguishes the two valleys. Explicitly, Hamiltonian for $ \tau = -1$ reads
	\begin{equation}
	    \label{eqn:hamiltonian_neg_xi}
	    H_{\tau=-1}=\left(
		\begin{array}{cccccc}
			V1& -p_{+}&0&0&0&0\\
			-p_{-}&V1&\gamma_{1}&0&0&0\\
			0&\gamma_{1}&V2& -p_{+}&0&0\\
			0&0&-p_{-}&V2&\gamma_{1}&0\\
			0&0&0&\gamma_{1}&V3& -p_{+}\\
			0&0&0&0&-p_{-}&V3\\				
		\end{array}
		\right).
	\end{equation}
	Similar to the bilayer case,\cite{landau-level-degeneracy-and-QHE-in-BLG} the Hamiltonian for valley $ \tau = -1$ can be transformed to the Hamiltonian for $ \tau = 1 $ if only the layer potential is interchanged by rearranging the basis. By adopting basis $ (-\psi_{B3},\psi_{A3},-\psi_{B2},\psi_{A2},-\psi_{B1},\psi_{A1}) $, the Hamiltonian in Eq.~(\ref{eqn:hamiltonian_neg_xi}) transforms to	
	\begin{equation}
	    \label{eqn:hamiltonian_neg_xi_trans}
	    \left(
		\begin{array}{cccccc}
			V3& p_{-}&0&0&0&0\\
			p_{+}&V3&-\gamma_{1}&0&0&0\\
			0&-\gamma_{1}&V2& p_{-}&0&0\\
			0&0&p_{+}&V2&-\gamma_{1}&0\\
			0&0&0&-\gamma_{1}&V1& p_{-}\\
			0&0&0&0&p_{+}&V1\\				
		\end{array}
		\right),
	\end{equation}
	which is identical to the Hamiltonian for $\tau =1$ except for an extra minus sign before $ \gamma_{1} $ and an interchange of layer 1 and 3. $\gamma_{1} $ always appears as squared through out the calculation so the minus sign has no effect on the energy levels. Hence we introduce $ \tau $ to account for the interchange of layer potentials, as adopted in Eq.~(\ref{eqts:potential_overlap_gap}) and thereafter. Similar convention is also adopted for BLG.\cite{EEInteractionInBLG,BoundStateAndBFieldInducedVS}
	

\section{\label{sec:expansionByLL}Expansion based on wave functions corresponding to Landau levels }
	
	The similarity between QD bound state spectrum and Landau levels in strong magnetic field regime implies a new way to handle this eigenvalue problem. The eigenfunction $\Psi_{i}$ under the QD potential can be expanded in the basis of LL states in polar coordinates $\psi_{n,m}^{p}$:
	\begin{equation}
	\Psi_{i}=\sum_{n,m}\sum_{p=1}^{6}a_{n,m}\psi_{n,m}^{p},
	\label{eqt:exp}
	\end{equation}
	where $i$ is the index for energy levels under a dot potential, $m$ is the angular momentum quantum number, $n$ is the principle quantum number and $p$ is the index of the six eigenvalues of the 6-by-6 matrix outside the dot. The confining potential well can be treated as a perturbation on a flat potential. The Hamiltonian is composed of two parts, $H=H_{0}+V(r,\theta)$. $H_{0}$ corresponds to the Hamiltonian outside the dot, while $V(r,\theta)$ corresponds to the potential perturbation inside the dot. $V(r,\theta)$ is diagonal and the diagonal terms are spacial dependent with zero values outside the dot. Denoting $\psi_{n,m}^{p}$ as $\ket{nmp}$, the matrix element Hamiltonian under this basis is
	\begin{eqnarray}
	\bra{nmp}H\ket{ijq}&&=\bra{nmp}H_{0}+V\ket{ijq}\nonumber\\
	&&=E_{np}\delta_{ni}\delta_{mj}\delta_{pq}+\bra{nmp}V\ket{ijq}.
	\label{eqt:Hmatele}
	\end{eqnarray}
	For explicitness, the LL states can be written as a combination of six components. 
	\begin{equation}
	\label{eqt:expLLr}
	\psi_{n,m}^{p}(r,\theta)=\sum_{s,p=1}^{6}c_{p,s}^{n,m}\Phi_{s}^{n,m}(r,\theta)\ket{s},
	\end{equation}
	where $\Phi_{s}^{n,m}$ is the spatial wave function and $s$ is the index for the spinor components (not the sign of $B$ as in Sec.~\ref{sec:bound_states} and $B>0$ in Sec.~\ref{sec:expansionByLL}). The space dependent potential on the basis of six components is a 6-by-6 matrix
	\begin{equation}
	V(r,\theta)=\sum_{s=1}^{6}	V_{s}(r,\theta)\ket{s}\bra{s},
	\end{equation}
	where the diagonal terms vanish outside the dot
	\begin{equation}
	V_{i}(r,\theta)=0~~(r>R)~~i=1,2 \dots 6.
	\end{equation}
	Thus the matrix element of the space dependent potential $V_{nmp,ijq}$ can be written as 

	\begin{eqnarray}
	&&V_{nmp,ijq}=\bra{nmp}V(r,\theta)\ket{ijq}\nonumber\\
	        &&=\iint \mathrm{d}A\left(\sum_{s,t=1}^{6}(c_{p,s}^{n,m}\Phi_{s}^{n,m})^{*}c_{q,t}^{i,j}\Phi_{t}^{i,j}\bra{s}V(r,\theta)\ket{t}\right)\nonumber\\
			&&=\iint_{r<R} \mathrm{d}A\left(\sum_{s=1}^{6}(c_{p,s}^{n,m}\Phi_{s}^{n,m})^{*}c_{q,s}^{i,j}\Phi_{s}^{i,j}V_{s}(r,\theta)\right).
	\label{eqt:Vmatele}
	\end{eqnarray}
	
	The above expression can be further simplified in the example mentioned in Sec.~\ref{sub:comparison}, whose potential is rotational symmetric. Eq.~(\ref{eqt:Vmatele}) becomes

	\begin{eqnarray}
		V_{nmp,ijq}&&=\iint_{r<R} \mathrm{d}A\left(\sum_{s=1}^{6}(c_{p,s}^{n,m}\Phi_{s}^{n,m})^{*}c_{q,s}^{i,j}\Phi_{s}^{i,j}V_{s}(r)\right)\nonumber\\
				&&=\delta_{m,j}C\int_{0}^{R}\mathrm{d}r \left(\sum_{s=1}^{6}(\varphi_{s}^{n,m})^{*}\varphi_{s}^{i,j} V_{s}(r)\right),
	    \label{eqt:reducedVmatele}
	\end{eqnarray}
	where $C=2\pi (c_{p,s}^{n,m})^{*}c_{q,s}^{i,j}/(N_{n,m}N_{i,j})$. The angular parts are orthogonal if $m\neq j$, hence $V_{nmp,ijq}=0$, implying that the hybridization only occurs between LL states of identical $m$.     
		
	When all the matrix elements have been calculated, the eigenvalues of this large matrix give the energy spectrum. If the base functions of all $n$'s and $m$'s are considered in Eq.~(\ref{eqt:exp}), the spectrum will be quite accurate, which is impossible. Nevertheless, only limited number of $n$ will suffice for the accuracy of a small range of energy levels from the perturbation theory, because the unperturbed energy generally varies with $n$ for each particular $p$ in Eq.~(\ref{eqt:exp}) . Without the perturbation potential, the degeneracy for each energy is infinite due to the different $m$'s. However, only limited number of $m$ will enter the perturbation process effectively because of the localization of the LL states.

	\subsection{\label{sub:LLr}Landau level wave functions in polar coordinates}

		Following the Hamiltonian in Eq.~(\ref{eqt:ham_m}) in polar coordinate and based on the spinor in Eq.~(\ref{eqt:eigenstate_psi}), for the homogeneous electric field $V_{i}(r)=V_{i}$ (constant), another form of the six-component wave function can be adopted to simplify the Hamiltonian:
		\begin{equation}
		\label{eqt:eigenstate_psi_n}
		\psi_{n,m}(r,\theta) = \frac{e^{im\theta}}{N_{n,m}\sqrt{r}}\left(
			\begin{matrix} 
			&c_{1}^{n,m}\varphi^{n,m}_{1}(r)&\\ 
			&c_{2}^{n,m}e^{i\theta}\varphi^{n,m}_{2}(r)&\\
			&c_{3}^{n,m}e^{i\theta}\varphi^{n,m}_{3}(r)&\\
			&c_{4}^{n,m}e^{2i\theta}\varphi^{n,m}_{4}(r)&\\
			&c_{5}^{n,m}e^{2i\theta}\varphi^{n,m}_{5}(r)&\\
			&c_{6}^{n,m}e^{3i\theta}\varphi^{n,m}_{6}(r)&\\
			\end{matrix} \right).
		\end{equation}
		In the above spinor, $n\in\mathbb{Z},m\in\mathbb{Z},n \ge -2,m<n$. $c_{s}^{n,m}$ is the coefficient of each component and $\sum_{s=1}^{6}|c_{s}^{n,m}|^{2}=1$. $N_{n,m}$ is the normalization factor and $N_{n,m}= \sqrt{\pi (n-1)!(n^{3}+5n^{2}+6n+1)/(n-m-1)!}$ for $n>0$. The specific forms of the components are as follows.
		\begin{eqnarray}
			\label{eqts:LLr}
			\varphi^{n,m}_{1} &=& \left\{
				\begin{array}{lcl}
					0								&	& {n<1},\\
					e^{-\xi^2/2}\xi^{m+1/2}L_{-1-m+n}^{m}(\xi^{2})	&	& {n\ge1},
				\end{array} \right. \nonumber\\
			\varphi^{n,m}_{2} = \varphi^{n,m}_{3} &=& \left\{
				\begin{array}{lcl}
					0									&	& {n<0},\\
					e^{-\xi^2/2}\xi^{m+3/2}L_{-1-m+n}^{m+1}(\xi^{2}) 	&	& {n\ge0},
				\end{array} \right. \nonumber\\
			\varphi^{n,m}_{4} = \varphi^{n,m}_{5} &=& \left\{
				\begin{array}{lcl}
					0									&	& {n=-2},\\
					e^{-\xi^2/2}\xi^{m+5/2}L_{-1-m+n}^{m+2}(\xi^{2}) 	&	& {n\ge-1},
				\end{array} \right.  \nonumber\\
			\varphi^{n,m}_{6} &=& e^{-\xi^2/2}\xi^{m+7/2}L_{-1-m+n}^{m+3}(\xi^{2}).	\nonumber\\
		\end{eqnarray}	
		In Eqs.~(\ref{eqts:LLr}), $\xi=r/(\sqrt{2}l_{B})$ and $L_{a}^{b}(x)$ is the generalized Laguerre polynomial. Under such basis, the eigen wave function can be expressed with the aforementioned coefficients $\Psi_{n,m}=(c_{1}^{n,m},c_{2}^{n,m},c_{3}^{n,m},c_{4}^{n,m},c_{5}^{n,m},c_{6}^{n,m})^{T}$. The Hamiltonian acting on this vector is also numeric, similar to Eq.~(\ref{eqt:ham_num}), but with $b_{i}$ replaced by another group of parameters $\beta_{i}$. The values of $\beta_{i}$ are as listed in Table.~\ref{tab:beta}
		\begin{table}
			\caption{\label{tab:beta}Values of $\beta_{i} $. }
			\begin{ruledtabular}
			\begin{tabular}{ccccc} 
				$\beta$ & $n \ge 1$ & $n=0$ & $n=-1$ & $n=-2$ \\
				\hline
				$\beta_{1} $ & $ 2n $ & $-$  & $-$  & $-$ \\
				$\beta_{2} $ & $ -2 $ & $-$  & $-$  & $-$ \\
				$\beta_{3} $ & $2n+2$ & $2$  & $-$  & $-$ \\
				$\beta_{4} $ & $-2$   & $-2$ & $-$  & $-$ \\
				$\beta_{5} $ & $2n+4$ & $4$  & $2$  & $-$ \\
				$\beta_{6} $ & $-2$   & $-2$ & $-2$ & $-$ \\	
			\end{tabular}
			\end{ruledtabular}
		\end{table}
		For $n=0,-1,-2$, the spinor always has zero components indicating the Hamiltonian should be reduced to a $(5-2|n|)$-by-$(5-2|n|)$ matrix, and the number of eigenenergies is also reduced.
		
	\subsection{\label{sub:locality} The locality of LL states}
	
		As introduced in Eq.~(\ref{eqt:expLLr}) and Eq.~(\ref{eqt:eigenstate_psi_n}), $\Phi_{s}^{n,m}$ can be decomposed into radius part and angular part. The reduced radial wave functions have a similar form, which is $e^{-\xi^2/2}\xi^{m+1/2+a}L_{-1-m+n}^{m+a}(\xi^{2})$, where the possible values of $n$ and $m$ are mentioned in Appendix.~\ref{sub:LLr} and $a=0,1, 2, 3$. These functions are all localized in space and exhibit similar dependency on $ m $. The spatial behavior of these functions are critical to the estimation of $V_{nmp,ijq}$ in Eq.~(\ref{eqt:Vmatele}) qualitatively.

\begin{figure}
		\includegraphics[width=3.4in]{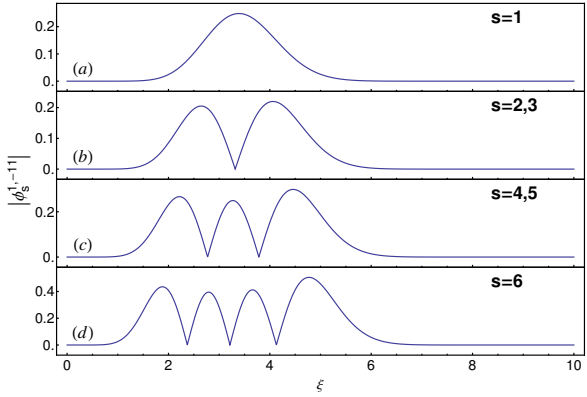}
		\caption{\label{fig:LLrmultis}Reduced radial wave functions of LLs in polar coordinates. In this figure, $n=1, m=-1$. The wave function with larger $s$ has equal or more zeros.}
		\end{figure}

		Figure~\ref{fig:LLrmultis} shows the LL reduced radial wave functions of the six components indexed by $s$ when $n=1, m=-1$. For $s=6$, the wave function has the most nodes and is most extensive. Thus it is reasonable to use the last component to estimate the spatial extension of the LL state.

		For larger $n$, the wave function has more zeros and is more extensive. For a normalized state, more extensive spatial extension indicates smaller probability amplitude within this range, which means the value of $V_{nmp,ijq}$ in Eq.~(\ref{eqt:Vmatele}) is small for the integral with different $n$ and $i$. So it can be expected that if $|n-i|$ is large enough, the perturbation matrix element itself will be small, not to mention that the energy differences further limit the perturbation effect.

\begin{figure}
		\includegraphics[width=3.4in]{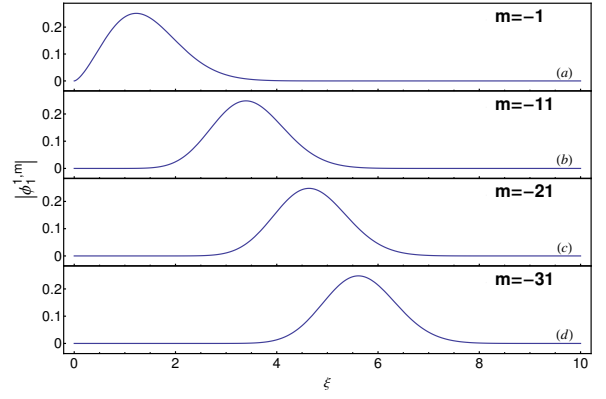}
		\caption{\label{fig:LLrmultim}Reduced radial wave functions of LLs in polar coordinates. $s = 1$, $n = 1$ and multiple value of $m$ are included. When $m\le0$, the wave packet is located further from the origin if $|m|$ is larger.}
		\end{figure}

		Figure~\ref{fig:LLrmultim} shows the LL reduced radial wave functions of the first components with different $m$'s for $n=1$. When $m\le0$, for larger $|m|$, the wave function is located further from the origin. This is the critical feature making the perturbation method possible here. Even though the original energy levels are degenerated due to the infinite number of $m$, the overlap between $\varphi_{s}^{n,m}$ and $\varphi_{s}^{n,j}$ is significantly small when $|m-j|$ is large enough. Given the magnetic field $B$ and the electric potential well, LL states with large $|m|$ are located outside the dot and can be effectively treated as eigen states, thus have little perturbation effects on other states. An upper limit of $m$ can be determined in this way, and only a finite number of LL states need to be recombined to form new states. If the potential inside the dot is flat, the energy levels can be acquired by solving the numeric Hamiltonian matrix with the diagonal elements replaced by the inner potential. In this case, hybridization only exists among the six components with the same $n$ and $m$ but different $p$'s.

\begin{figure}
		\includegraphics[width=3.4in]{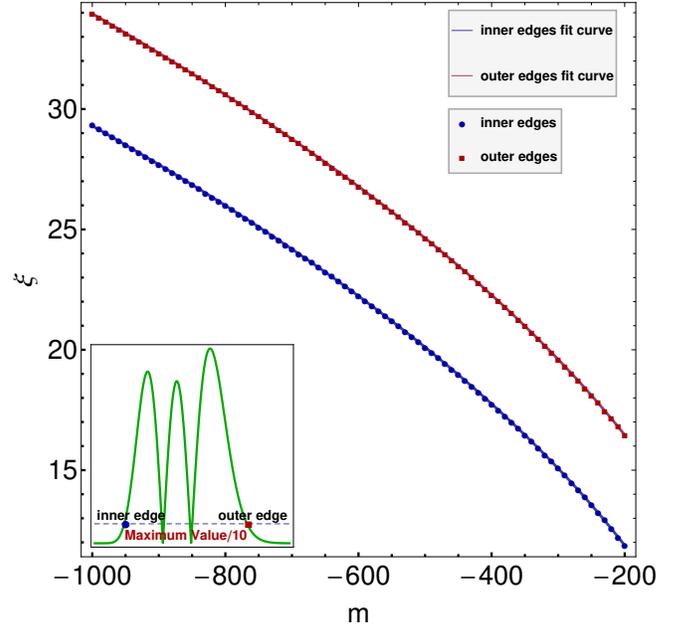}
		\caption{\label{fig:EdgeFit} Edges of the reduced radial LL wave functions and corresponding nonlinear fit curves. The blue dots are the inner edges of the last component of the reduced radial wave function for $n=0$, and the red squares are the outer edges. The edges are defined as the inner and outer most positions where the norm of the wave function has fallen to one tenth of its maximum value, which are shown in the inset. For $m<0$, the trends of the edges versus $m$ are close to $\sqrt{|m|}$ and can be perfectly fitted by $a+\sqrt{b|m|+c}$. }
		\end{figure}

		In Fig.~\ref{fig:EdgeFit} the spatial extensions of the reduced radial wave functions of different $m$'s for $n=0$ are calculated and displayed as a function of $m$. The spatial extension of the given wave function is defined by the inner and outer edge. The edges are defined as the inner and outer most positions where the norm of the wave function has fallen to one tenth of its maximum value, and the edges can be perfectly fitted by $a+\sqrt{b|m|+c}$.

		 Notice here the spatial extension is defined by $\xi$, which is $r/(\sqrt{2}l_{B})$. As $B$ increases, the range of $r$ will be shrunk towards the origin. For a step potential well with certain $m$ and given magnetic field $B$, if the dot edge is within the spatial extension, all such $B$'s compose the transition range for energy level with this $m$.

		In this way, once the transition range is known, the tendency of the energy spectrum can be roughly predicted. Instead of computing the integrals and solving for the eigen values of the Hamiltonian in the perturbation process, the validity of the method in Sec.~\ref{sub:analytic_solution} is tested by focusing on the locality of the LL states to predict the transition range. The ability of such prediction is a significant advantage of the expansion method. It is time-consuming to calculate the eigenenergies one by one, but it is much easier, by using simple math, to obtain the spatial extension of a LL state and compare it with the dot edge.
		
	\section{\label{sec:note} Note added in proof}
	    When preparing this paper for publication, we became
        aware of the related paper by M. Mirzakhani \textit{et al.}\cite{PhysRevB.94.165423} These two papers differ in boundary conditions and calculation methods. Under a finite step well potential, we show the transition of bound state levels between different groups of LLs and relate it to the positions of LL states through perturbation theory.
	

\nocite{*}

\bibliographystyle{apsrev4-1}
\bibliography{TLGtheory}

\end{document}